\documentclass[10pt]{blois07}
\usepackage{graphicx}
\usepackage{cite,./mcite}

\begin{document}
\title{Extended Air Shower Simulations Based on EPOS}
\author{Klaus WERNER$^1$\protect\footnote{\ talk presented at EDS07}\  and
 Tanguy PIEROG$^2$}
\institute{$^1$ SUBATECH, University of Nantes -- IN2P3/CNRS-- EMN, 
  Nantes, France       \\ 
$^2$Forschungszentrum Karlsruhe, Institut f\"ur Kernphysik,
    Karlsruhe, Germany}
\maketitle
\begin{abstract}
 We discuss air shower simulations based on the EPOS hadronic interaction model.
 A remarkable feature is the fact that the number of produced muons is
 considerably larger compared to other interaction models. We show that this is
 due to an improved treatment of baryon-antibaryon production.
\end{abstract}

\section{Introduction}

Air shower simulations are a very powerful tool to interpret ground
based cosmic ray experiments. However, most simulations are still
based on hadronic interaction models being more than 15 years old.
Much has been learned since, in particular due to new data available
from the SPS and RHIC accelerators. 

In this paper, we discuss air shower simulations based on EPOS, the
latter one being a hadronic interaction model, which does very well
compared to RHIC data~\cite{Bellwied,Abelev}, and also all other
available data from high energy particle physic experiments (ISR,CDF
and especially SPS experiments at CERN). 

EPOS is a consistent quantum mechanical multiple scattering approach
based on partons and strings~\cite{nexus}, where cross sections
and the particle production are calculated consistently, taking into
account energy conservation in both cases (unlike other models where
energy conservation is not considered for cross section calculations~\cite{hladik}).
A special feature is the explicit treatment of projectile and target
remnants, leading to a very good description of baryon and antibaryon
production as measured in proton-proton collisions at 158~GeV at
CERN~\cite{nex-bar}. Nuclear effects related to CRONIN transverse
momentum broadening, parton saturation, and screening have been introduced
into EPOS~\cite{splitting}. Furthermore, high density effects leading
to collective behavior in heavy ion collisions are also taken into
account~\cite{corona}.

\section{EPOS Basics}

\begin{figure}
\begin{center}(a)\includegraphics[%
  scale=0.35]{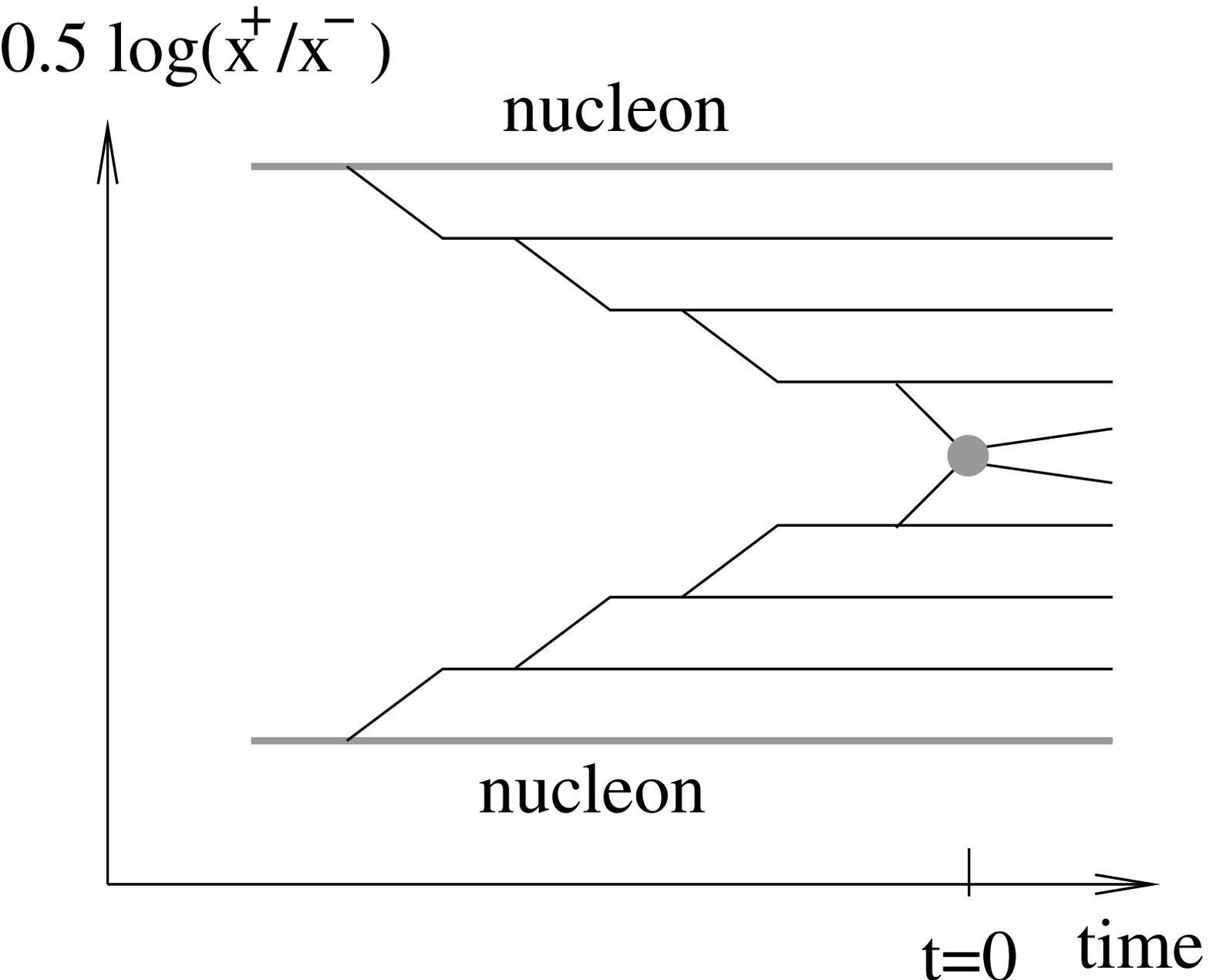}$\qquad$(b)~\includegraphics[%
  scale=0.6]{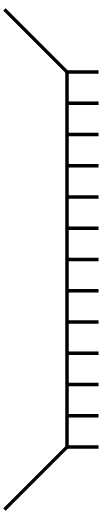}$\qquad$(c)~\includegraphics[%
  scale=0.55]{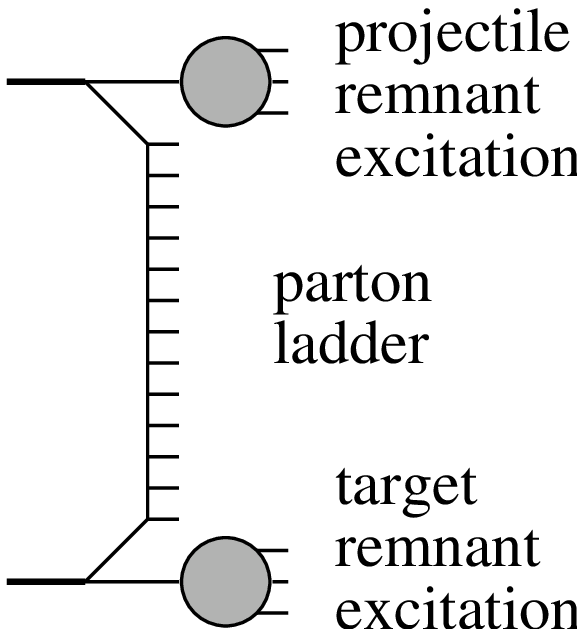}\end{center}

\caption{(a) Elementary parton-parton scattering: the hard scattering in the
middle is preceded by parton emissions (initial state space-like cascade).
(b) Symbolic parton ladder, representing the structure shown left.
(c) The complete picture, including remnants. The remnants are an
important source of particle production at RHIC energies.\label{ladder} }
\end{figure}

One may consider the simple parton model to be the basis of hadron-hadron
interaction models at high energies. It is well known that the inclusive
cross section is given as a convolution of two parton distribution
functions with an elementary parton-parton interaction cross section.
The latter one is obtained from perturbative QCD, the parton distributions
are deduced from deep inelastic scattering. Although these distributions
are taken as black boxes, one should not forget that they represent
a dynamical process, namely the successive emission of partons (initial
state space-like cascade), as shown in fig. \ref{ladder}(a). We refer
to this whole structure as {}``parton ladder'', with a corresponding
simple symbol as shown in fig. \ref{ladder}(b), to simplify further
discussion.

Actually our {}``parton ladder'' is meant to contain two parts \cite{nexus}:
the hard one, as discussed above, and a soft one, which is a purely
phenomenological object, parametrized in Regge pole fashion.

Still the picture is not complete, since so far we just considered
two interacting partons, one from the projectile and one from the
target. These partons leave behind a projectile and target remnant,
colored, so it is more complicated than simply projectile/target deceleration.
One may simply consider the remnants to be diquarks, providing a string
end, but this simple picture seems to be excluded from strange antibaryon
results at the SPS \cite{sbaryons}.

We therefore adopt the following picture, as indicated in fig. \ref{ladder}(c):
not only a quark, but a two-fold object takes directly part in the
interaction, being a quark-antiquark, or a quark-diquark, leaving
behind a colorless remnant, which is, however, in general excited
(off-shell). So we have finally three {}``objects'', all being white:
the two off-shell remnants, and the parton ladders between the two
active {}``partons'' on either side (by {}``parton'' we mean quark,
antiquark, diquark, or antidiquark). We showed in ref. \cite{nex-bar}
that the {}``three object picture'' as discussed in this paper can
solve the {}``multi-strange baryon problem'' of ref. \cite{sbaryons}.

Even inclusive measurements require often more information than just
inclusive cross sections, for example via trigger conditions. Anyhow,
for detailed comparisons we need an event generator, which obviously
requires information about exclusive cross sections (the widely used
pQCD generators are not event generators in this sense, they are generators
of inclusive spectra, and a Monte Carlo event is not a physical event).
This problem is known since many years, the solution is Gribov's multiple
scattering theory, employed since by many authors. This formulation
is equivalent to using the eikonal formula to obtain exclusive cross
sections from the knowledge of the inclusive one.

We indicated recently inconsistencies in this approach, proposing
an {}``energy conserving multiple scattering treatment'' \cite{nexus}.
The main idea is simple: in case of multiple scattering, when it comes
to calculating partial cross sections for double, triple ... scattering,
one has to explicitly care about the fact that the total energy has
to be shared among the individual elementary interactions.

A consistent quantum mechanical formulation of the multiple scattering
requires not only the consideration of the (open) parton ladders,
discussed so far, but also of closed ladders, representing elastic
scattering.
The closed ladders do not contribute to particle production, but they
are crucial since they affect substantially the calculations of partial
cross sections. Actually, the closed ladders simply lead to large
numbers of interfering contributions for the same final state, all
of which have to be summed up to obtain the corresponding partial
cross sections. It is a unique feature of our approach to consider
explicitly energy-momentum sharing at this level (the {}``E'' in
the name EPOS).

\section{Splitting of Parton Ladders}

Let us consider very asymmetric nucleus-nucleus collisions, like proton-nucleus
or deuteron-nucleus. The formalism developed earlier for $pp$ can
be generalized to these nuclear collisions, as long as one assumes
that a projectile parton always interacts with exactly one parton
on the other side, elastically or inelastically (realized via closed
or open parton ladders). We employ the same techniques as already
developed in the previous section. The calculations are complicated
and require sophisticated numerical techniques, but they can be done.

\begin{figure}
\begin{center}\includegraphics[%
  scale=0.55]{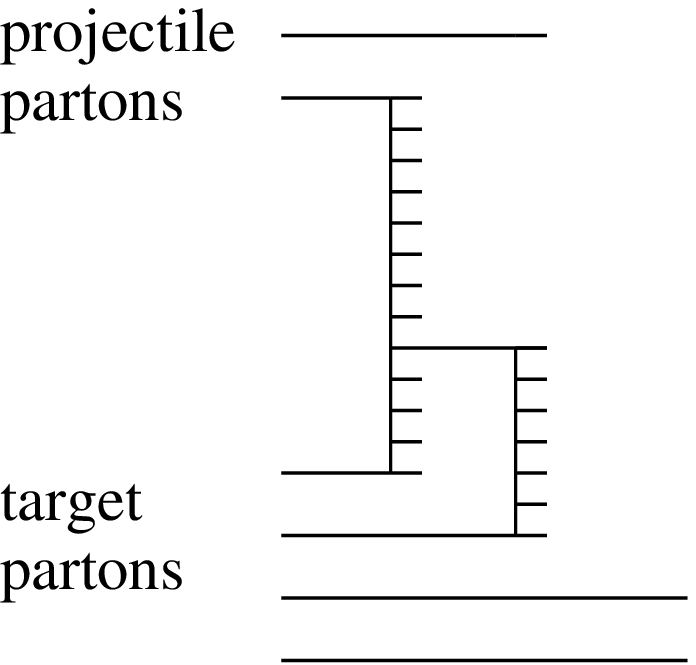}$\qquad$\includegraphics[%
  scale=0.55]{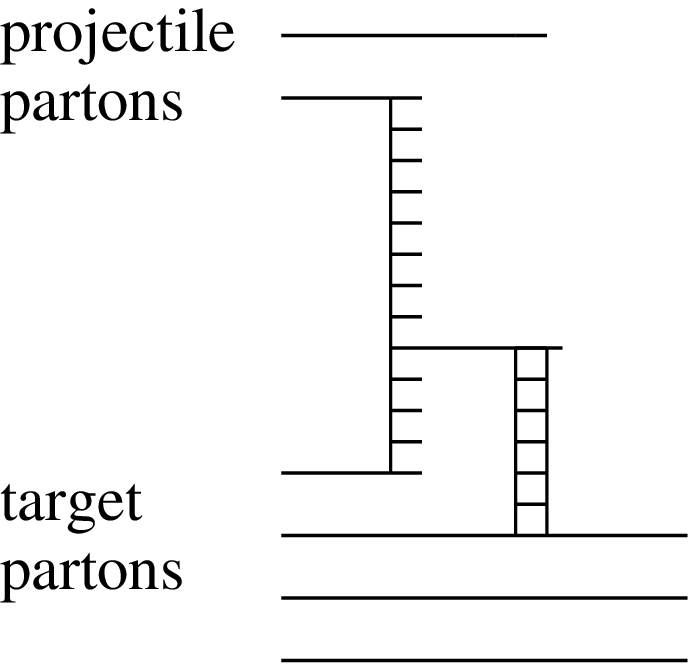}$\,$\end{center}

\caption{Inelastic and elastic {}``rescattering'' of a parton from the parton
ladder with a second target parton. We talk about (inelastic and elastic)
splitting of a parton ladder. \label{split}}
\end{figure}

In case of protons (or deuterons) colliding with heavy nuclei (like
gold), there is a complication, which has to be taken into account:
suppose an inelastic interaction involving an open parton ladder,
between a projectile and some target parton. The fact that these two
partons interact implies that they are close in impact parameter (transverse
coordinate). Since we have a heavy target, there are many target partons
available, and among those there is a big chance to find one which
is as well close in impact parameter to the two interacting partons.
In this case it may be quite probable that a parton from the ladder
interacts with this second target parton, inelastically or elastically,
as shown in fig. \ref{split}.

The main effect of elastic splittingis suppression of small light cone momenta,
which agrees qualitatively with the concept of
saturation. But this is only a part of the whole story, several other
aspects have to be considered\cite{splitting}. 
Consider the example shown in figure \ref{split}(left). 
In the upper part, there
is only an ordinary parton ladder, so we expect ``normal'' hadronization.
In the lower part, we have two ladders in parallel, which are in addition
close in space, since they have a common upper end, and the lower
ends are partons close in impact parameter, so the hadronization of
the two ladders is certainly not independent, we expect some kind
of a {}``collective'' hadronization of two interacting ladders.
Here, we only considered the most simple situation, one may also imagine
three or more close ladders, hadronizing collectively.

The strength of the effects due to parton ladder splitting will depend
on the target mass, via the number $Z$ of partons available for additional
legs. The number $Z$ of available partons will also increase with
energy, so at high enough energy the above-mentioned effects can already
happen in $pp$ collisions.

A quantitative discussion how the above-mentionned effects are realized
may be found in \cite{splitting}.

\section{Air Shower Simulations}

In the following, we discuss air shower simulations, based on the shower
programs CORSIKA\cite{corsika} or CONEX\cite{conex0,conex}, using
EPOS or QGSJET II-3\cite{qgsjetII} (as a reference) as interaction model. 

\begin{figure}[h]
\begin{center}\includegraphics[width=0.48\textwidth]{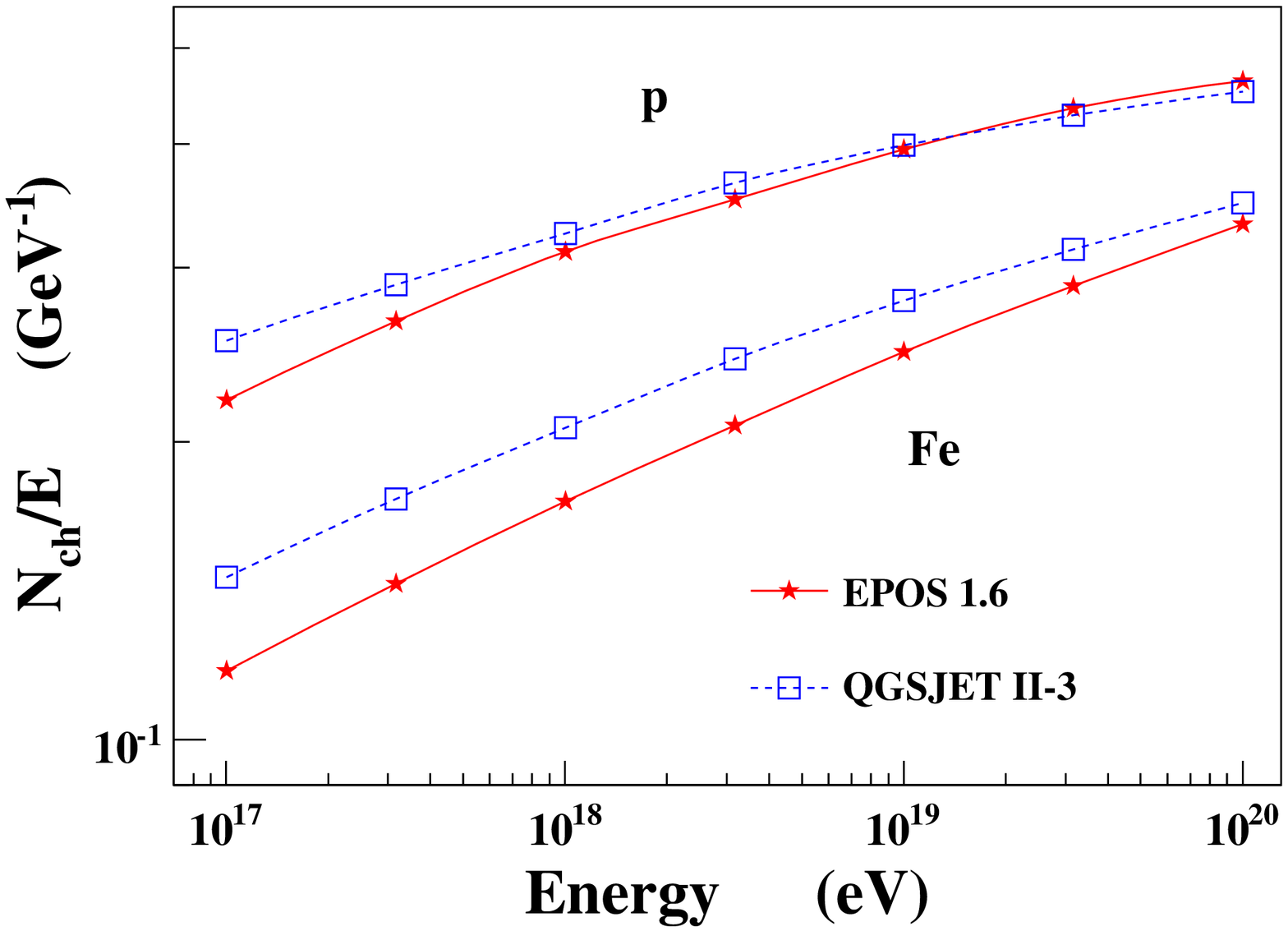}
\includegraphics[width=0.48\textwidth]{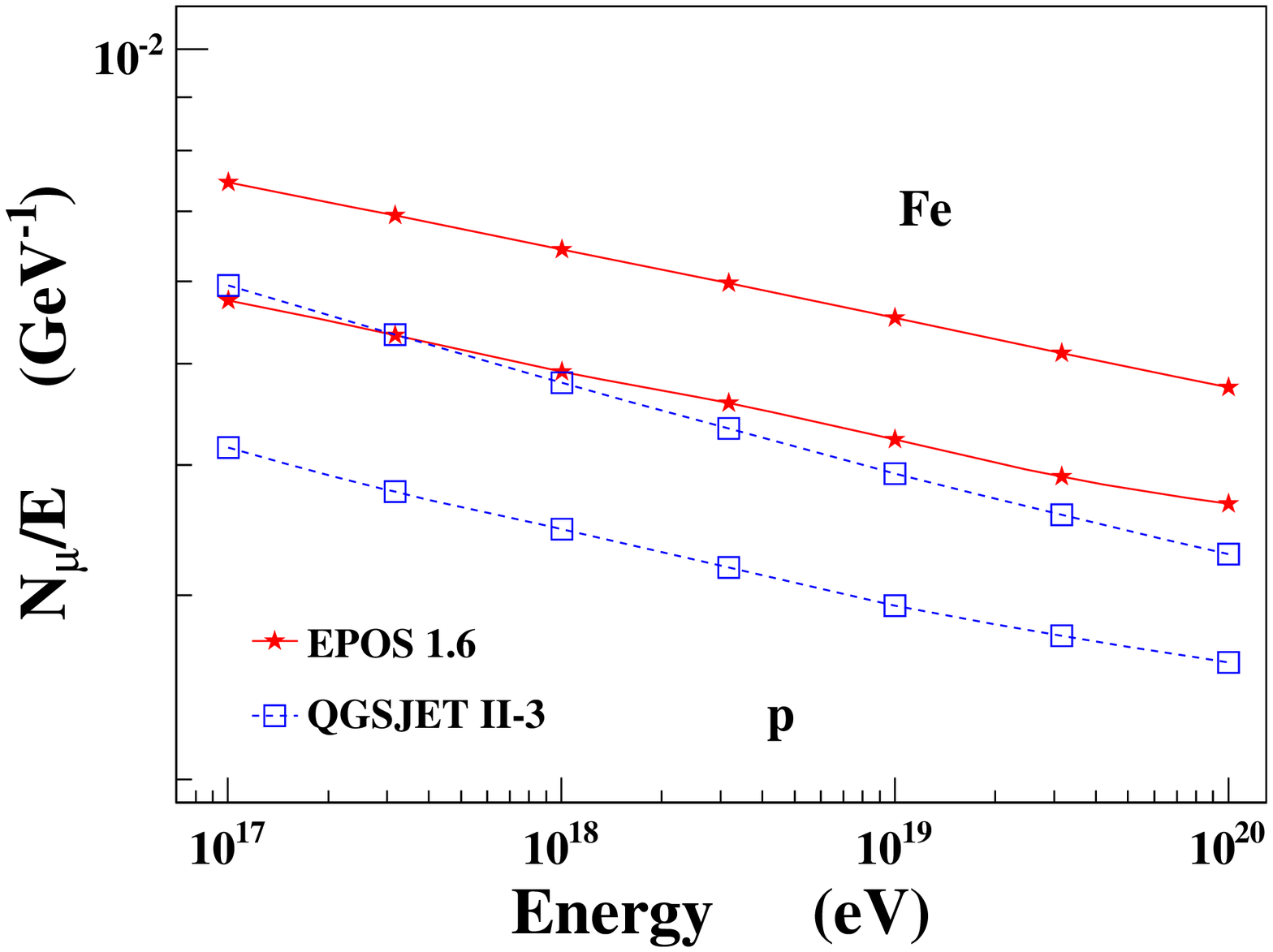}\end{center}
\caption{Total number of charged particles (left plot) and muons (right plot)
 at ground divided by the primary 
energy as a function of the primary energy for proton and iron induced 
shower using EPOS (full lines) and QGSJET~II-3 (dotted lines) as high 
energy hadronic interaction model.\label{fig-ch}}
\end{figure}

Air shower simulations are very important to analyze the two most common types 
of high energy cosmic ray experiments: fluorescence telescopes and surface detectors. In the first ones, one observes directly 
the  longitudinal shower development,
from which the energy and the depth of shower maximum 
$\mathrm{X}_{\mathrm{max}}$ can be
extracted. Comparing the latter with models allows us to have informations on
the mass of the primary. EPOS results concerning 
$\mathrm{X}_{\mathrm{max}}$ are in good agreement with former models and
experimental data.

Concerning particles measured at ground by air shower experiment, the situation
is quite different. Whereas the number of
charged particles is very similar for EPOS and QGSJET~II-3 (see fig.
\ref{fig-ch}), EPOS produces a much higher muon 
flux, in particular at high energy. At $10^{20}\,$eV
 EPOS is more than 40\% higher and gives even more
muons with a primary proton than QGSJET~II-3 for iron induced showers.

The muon excess from EPOS compared to other models will affect all experimental
observables depending on simulated muon results. In
the case of the Pierre Auger obervatory (PAO), this will affect mostly the results on inclined
showers, for which the electromagnetic component is negligible at ground. It is
interesting  to notice that the PAO claims  a possible 
lack of muons in air showers simulated with current hadronic interaction 
models.

\section{The origin of the increased muon production}

During the hadronic air shower development, the energy is
shared between neutral pions which convert their energy into the 
electromagnetic component of the shower, and charged hadrons which 
continue the hadronic cascade producing muons. 
The ratio  of the two (referred to as $R$) is a measure of the muon 
production.

\begin{figure}\begin{center}
\includegraphics[scale=0.7]{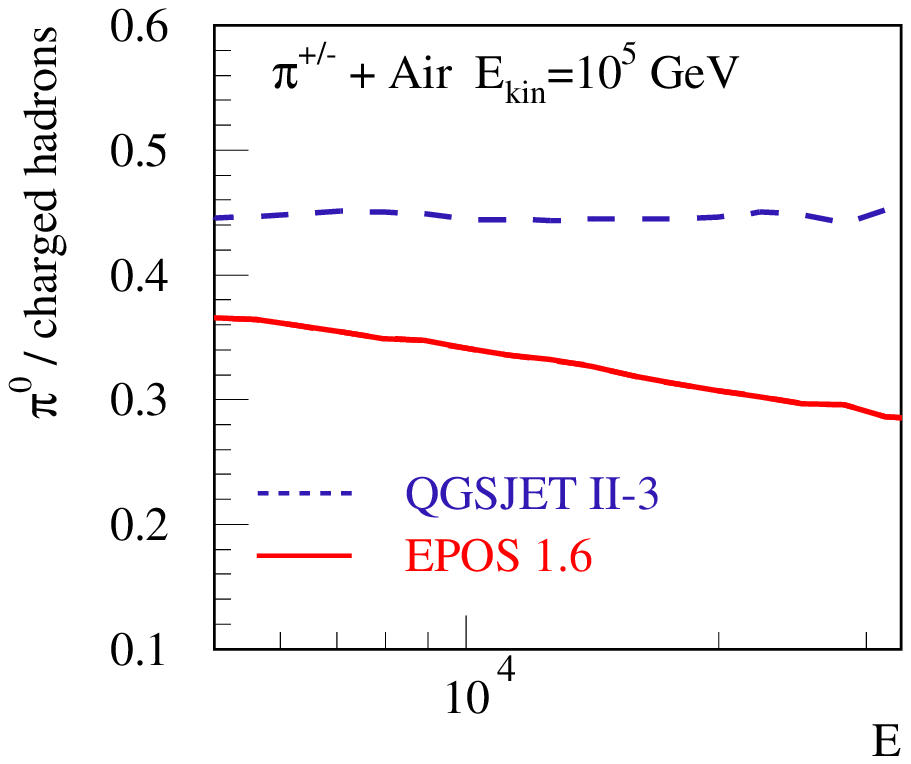}
%\end{figure}\begin{figure}
\includegraphics[scale=0.7]{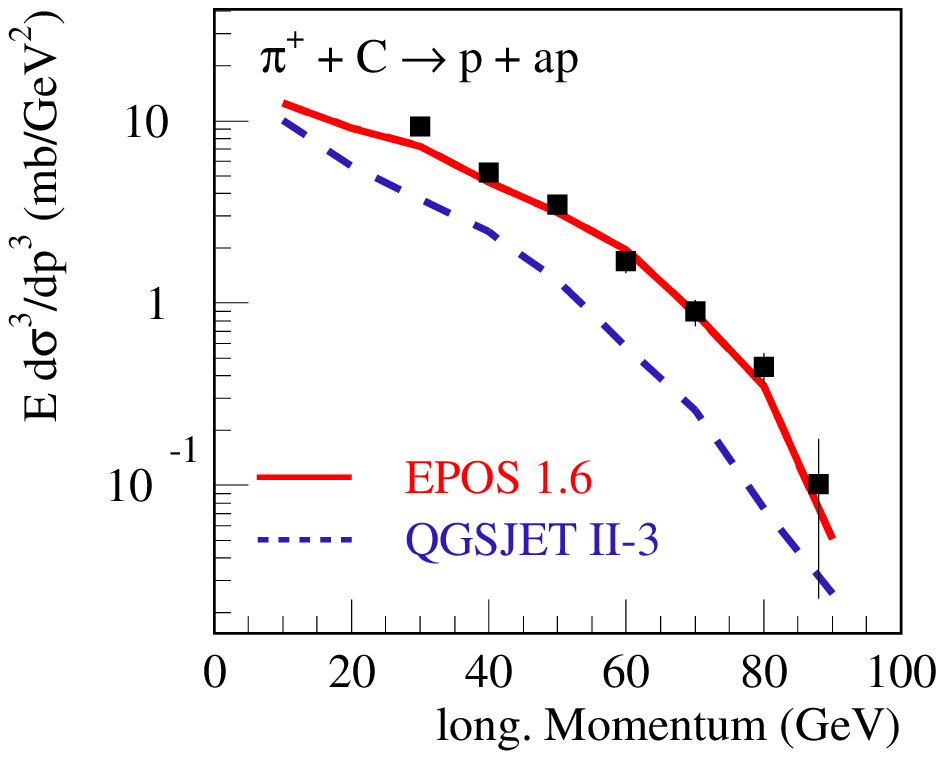}
\end{center}
\caption{
Left: Ratio of the number of $\pi^0$ over the number of charged 
particles as a function of the energy of the secondary particles at 
$10^{5}$~GeV kinetic energy with EPOS 
(full line) or QGSJET~II-3 (dashed line) in pion-air.
Right: Longitudinal momentum distributions of protons
in pion carbon collisions at 100 GeV from EPOS (full) and QGSJET~II-3
(dashed) compared to data.}

\label{fig-ratio}
\end{figure}

Comparing EPOS to other models, this ratio $R$ of neutral pions to charged 
hadrons
produced in individual hadronic interactions is significantly lower, 
especially for pi-air reactions, as seen in fig. \ref{fig-ratio}(left).
This will   increase the  muon production, as discussed above. 

Furthermore, the reduced ratio $R$ is partly
due to an enhanced baryon production, as shown in fig. \ref{fig-ratio}(right) (data from \cite{barton}).
This  will increase the number of baryon initiated sub-showers.
 Since the ratio $R$  is much softer 
in case of proton-air interactions compared to pion-air interactions, this will even more reduce $R$, 
providing a significant additional source of muons.

\section{Summary}

EPOS is a new interaction model constructed on a solid theoretical
basis. It has been tested very carefully against all existing hadron-hadron
and hadron nucleus data, also those usually not considered important
for cosmic rays. In air shower simulations, EPOS provides more muons
than other models, which was found to be linked to an increased baryon
production.

\begin{footnotesize}

\end{footnotesize}


\begin{thebibliography}{10}
\bibitem{Bellwied}R.~Bellwied {[}STAR Collaboration{]}, arXiv:nucl-ex/0511006. 
\bibitem{Abelev}B.~Abelev {[}STAR Collaboration{]}, arXiv:nucl-ex/0607033. 
\bibitem{nexus}H. J. Drescher, M. Hladik, S. Ostapchenko, T. Pierog, and K. Werner,
Phys. Rept. 350, 93, 2001 
\bibitem{hladik}M.~Hladik, H. J. Drescher, S. Ostapchenko, T. Pierog, 
  and K. Werner \textit{et al.}, Phys.\ Rev.\ Lett.\ {} \textbf{86},
3506 (2001), arXiv:hep-ph/0102194. 
\bibitem{nex-bar}F.M. Liu, J.Aichelin, M.Bleicher, H.J. Drescher, S. Ostapchenko, T.
Pierog, and K. Werner, Phys. Rev. D67, 034011, 2003 
\bibitem{splitting}K.~Werner, F.~M.~Liu, and T.~Pierog, Phys.\ Rev.\ C \textbf{74}
(2006) 044902. 
\bibitem{corona}K.~Werner, Phys. Rev. Lett. 98, 152301 (2007), arXiv: 0704.1270. 
\bibitem{sbaryons}M. Bleicher, F. M. Liu, A. Keränen, J. Aichelin, S.A. Bass, F. Becattini,
K. Redlich, and K. Werner, Phys.Rev.Lett.88, 202501, 2002. 
%\bibitem{mia}T. Abu-Zayyad et al. (HiRes-MIA Collab.) ,Astrophys. J. \textbf{557},
%686 (2001), arXiv: astro-ph/0010652
%\bibitem{hires}D.~R.~Bergman (HiRes Collaboration, Nucl. Phys. Proc. Suppl. \textbf{136},
%40 (2004), arXiv: astro-ph/0407244{]}. 
%\bibitem{qgsjet01}N.N.~Kalmykov, S.S.~Ostapchenko, A.I.~Pavlov, Nucl.~Phys.~B (Proc.~Suppl.)
%\textbf{52}, 17 (1997). 
\bibitem{corsika}D.~Heck \textit{et al.}, Report \textbf{FZKA 6019}, and D.~Heck,
J.~Knapp, Report \textbf{FZKA 6097}, Forschungszentrum Karls\-ruhe
(1998). 
\bibitem{conex0}G. Bossard, H.J. Drescher, N.N. Kalmykov, S. Ostapchenko, A.I. Pavlov,
T. Pierog, E.A. Vishnevskaya, and K. Werner, Phys.Rev. \textbf{D63},
054030, (2001)
\bibitem{conex}T.~Bergmann \textit{et al.}, arXiv:astro-ph/0606564. 
\bibitem{qgsjetII}S.Ostapchenko, Phys. Rev.D74, 014026 (2006)
bibitem{engel}R. Engel, T.K. Gaisser, P. Lipari, T. Stanev, Proc. 26th Int. Cosmic
Ray Conf., Salt Lake City, 415 (1999).
\bibitem{barton}D. S. Barton et al., Phys. Rev. D \textbf{27} (1983) 2580. 
\end{thebibliography}
\end{document}